\documentclass[a4paper,11pt]{article}
    \usepackage{cite}
    \usepackage{amsmath,amssymb}
    \usepackage{bm}
    \usepackage{mathrsfs}
    \usepackage[dvipdfm]{graphicx}
    \usepackage[abs]{overpic}
    \usepackage{pict2e}
    \newcommand{\be}{\begin{equation}}
    \newcommand{\ee}{\end{equation}}
    \newcommand{\nbe}{\begin{equation*}}
    \newcommand{\nee}{\end{equation*}}
    \newcommand{\regge}{\alpha'}
    \newcommand{\dgravc}{\kappa_d^2}
    \newcommand{\GBc}{\frac{\regge}{8}}
    \newcommand{\GBinv}{\mc{R}^2_\mathrm{GB}}
    \newcommand{\mc}{\mathcal}

    \newcommand{\pd}{\partial}
    \newcommand{\defeq}{\stackrel{\mathrm{def}}{=}}
    \newcommand{\half}{\frac{1}{2}}
    \newcommand{\EGB}{\text{Einstein-Gauss-Bonnet}}
    \newcommand{\DEGB}{\text{Dilatonic Einstein-Gauss-Bonnet}}
    \usepackage{layout}

\begin{document}
\title{\textbf{More on the dilatonic Einstein-Gauss-Bonnet gravity} \\ \phantom{DEGB}}
\author{\Large{Masao Iihoshi}\footnote{Electronic address: iihoshi@kiso.phys.se.tmu.ac.jp}
                \\ \\
                \emph{Department of Physics, Tokyo Metropolitan University,} \\
    \emph{Hachioji, Tokyo 192-0397, Japan}}
\date{\empty} 
\maketitle

\begin{abstract}
$\EGB$ gravity coupled to a dynamical dilaton is examined from the viewpoint of Einstein's equivalence principle.
We point out that the usual frame change that applies to the action without curvature correction does not cure the problem of
nonminimal couplings by the dynamical nature of a dilaton field.
Thus a modification of the Einstein frame is required.
It is proposed that the kinetic term of a dilaton should be brought to a canonical form, which completely fixes
the additional terms associated with the frame transformation.
\end{abstract}

\section{Introduction}
It is fairly natural to regard General Relativity (GR) as the low energy effective theory of gravity
which applies to physical phenomena well below the Planck scale, and is derived from quantum theory of gravity
that is expected to explain all known interactions quantum-theoretically.
In superstring/M-theory\cite{GSWP}, which is the major candidate for the consistent quantum theory of gravity at present,
GR emerges from the consideration of tree-level scattering amplitudes of string NS-NS fields
(or renormalization of a string world-sheet nonlinear sigma model).

It is also well known that string perturbation theory modifies GR,
which is known as $\regge$-corrections to the Einstein-Hilbert action, with $\regge$ being the Regge slope
(see Ref. \cite{GSWP} and references therein).\footnote{There is another modification known as a loop expansion
in powers of the string coupling constant $g_\mathrm{s}$, which we do not discuss here.}
In heterotic string we are interested in this paper, the modification starts with the first order in $\regge$
that defines the $\DEGB$ (DEGB) theory of gravity, i.e.
$\EGB$ (EGB) gravity with a dynamical dilaton field.\footnote{In fact heterotic string itself does not predict EGB gravity, involving
the \emph{particular} combination of quadratic terms in the spacetime curvature.
Because of the invariance of the string scattering amplitudes by local redefinitions of metric, there are ambiguities in curvature corrections
in the low energy effective action; see, e.g. Ref.\cite{GSWP}.
See Ref.\cite{Zwiebach} for the physical arguments leading to the EGB combination.}

Because a dilaton field $\phi$ determines the string coupling constant $g_\mathrm{s}$ through the vacuum expectation value of $e^{\phi}$,
it is quite important to include the dynamics of a dilaton.
For example, DEGB gravity (and its phenomenological generalization) is examined and reviewed in
Refs.\cite{BD86,KMRTW,TYM,NOS05,NOS06,LN,BGO,NT,CENO08,OT,MOS,GS,CGOO,NO10}.

As is well known, a string perturbation itself recaptures a gravitational action that is written in terms of the so-called string frame,
in which the coupling of gravitational field to a dilaton field is nonminimal.
Accordingly, in the ordinary procedure, one makes a conformal transformation to the Einstein frame action
in which gravitational couplings to another fields are minimal, in order to give the theory a physical interpretation.
When applying the above procedure to the DEGB gravity, the story becomes complicated.
Because of the coupling of the Gauss-Bonnet (GB) term to a dilaton, some additional coupling terms appear when one changes a frame,
and they are often simply discarded in the literature.\footnote{In Ref. \cite{MOS} the effect of the additional terms
on a black hole solution is studied.}

In this note the influence of those additional couplings is discussed from the viewpoint of Einstein's Equivalence Principle (EEP).
In the next section we observe that such additional terms obviously give a noncanonical kinetic term for a dilaton and break the EEP,
and hence the definition of the Einstein frame metric receives a modification to restore the EEP.
It is also argued that in almost all cases the dilaton kinetic term can be safely brought to a canonical form, by a dilaton field redefinition.
In the final section we summarize our results.

\section{Analysis}
The action of $d$-dimensional DEGB gravity in string frame is given by
\begin{gather}
I\bigl[g^\mathrm{st},\phi\bigr]
= \frac{1}{2\dgravc}\int d^dx\sqrt{-g^{\mathrm{st}}} e^{-2\phi}\left[ R^\text{(st)} + 4g^{\mathrm{st}\mu\nu}\pd_\mu\phi\pd_\nu\phi
                                                                                                    + \GBc\GBinv\bigl(g^\mathrm{st}\bigr) \right]; \notag \\
\GBinv \defeq R_{\mu\nu\lambda\rho}R^{\mu\nu\lambda\rho} - 4R_{\mu\nu}R^{\mu\nu} + R^2,\label{eqn:GBstring}
\end{gather}
where the superscript ``st'' stands for the quantities that are written with the string frame metric $g_{\mu\nu}^\mathrm{st}$.
By the following change:
\be
g_{\mu\nu}^\mathrm{st} = e^{\gamma'\kappa_d\varphi}g_{\mu\nu},
\;\text{ where }\; \varphi = \frac{\gamma'}{\kappa_d}\phi \;\text{ with }\;\gamma' = \frac{2}{\sqrt{d-2}}, \label{eqn:1st-trans}
\ee
the action \eqref{eqn:GBstring} reduces to \cite{MOS}
\begin{align}
I[g,\varphi] = \frac{1}{2\dgravc}\int d^dx\sqrt{-g}\biggl[ &R - \dgravc(\pd\varphi)^2
                                                                                   + {\GBc}e^{-\gamma'\kappa_d\varphi}\biggl\{ \GBinv(g)
                                                                                   + (\gamma'\kappa_d)^2d(d-3)G^{\mu\nu}\pd_\mu\varphi\pd_\nu\varphi \notag\\
                                                                                   &+ {\half}(\gamma'\kappa_d)^3(d-1)_3\nabla^2\varphi(\pd\varphi)^2
                                                                                   + \frac{1}{16}(\gamma'\kappa_d)^4(d-1)_4\left\{(\pd\varphi)^2\right\}^2 \biggr\}\biggr] \notag\\
                                                                                   &+ \text{ (surface terms)}, \label{eqn:GBfalseeinstein}
\end{align}
where $(\pd\varphi)^2 = g^{\mu\nu}\pd_\mu\varphi\pd_\nu\varphi$ and $\nabla^2\varphi = g^{\mu\nu}\nabla_\mu\pd_\nu\varphi$.
We have used the shorthanded notation $(d-m)_n = (d-m)(d-(m+1))\dots(d-n)$ with $n>m$.
In this paper we focus on a local variation, so we ignore surface terms in the action from now on.

When the action \eqref{eqn:GBfalseeinstein} is applied to a four-dimensional theory,
we are faced with the breakdown of EEP by the following two terms:
\nbe
e^{-\gamma'\kappa_d\varphi}\GBinv(g) \,\text{ and }\, e^{-\gamma'\kappa_d\varphi}G^{\mu\nu}\pd_\mu\varphi\pd_\nu\varphi.
\nee
As is mentioned in Introduction, the first term is the big appeal of perturbative strings
that brought the \emph{quadratic} terms in the curvature.
So it may play an important role in the very early stage of the Universe,
which we expect.\footnote{The signatures of a coupling between a scalar field (inflaton)
and the GB term on inflationary observables are studied in Ref.\cite{GS}.}
In the present low-curvature (almost flat) Universe, it can be safely ignored.

However, the presence of the second term could be dangerous, since it occurs with the \emph{linear} factor in the spacetime curvature.
As an example, let us consider the inflating flat FRW spacetime.
In this background we have the Ricci scalar $R \simeq 12H^2$ with $H$ being the expansion parameter whose typical value is of the order of
$10^{15}\mathrm{GeV}$\cite{LL}. The latter coupling may give a significant imprint on inflationary observables.

In fact, one may advocate that this coupling would cause no significant problem,
since $\regge$ gives the string length scale $\ell_\mathrm{s}$ by $\ell_\mathrm{s} = \sqrt{\regge}$
and so is very small, when compared to a typical scale of low energy effective theory.
However, then, this nonminimal coupling would suffer from a constraint by cosmological observations \cite{TG}.
It seems to be natural to demand that the EEP is valid \emph{exactly}.
Originally the Einstein frame is introduced to recover GR (i.e. the EEP). It would be appropriate to extend the definition of Einstein frame,
regardless of the value of $\regge$:
\emph{Einstein frame should be a frame in which the EEP exactly holds up to the first order in the curvature.}

According to our proposal, the transformation \eqref{eqn:1st-trans} does not define the Einstein frame
and hence Eq.\eqref{eqn:GBfalseeinstein} is not the Einstein frame action.
We find the genuine Einstein frame below.

In addition to \eqref{eqn:1st-trans}, we perform the following field redefinition (shift of the metric):
\be
g_{\mu\nu} \rightarrow g^\mathrm{new}_{\mu\nu} = g_{\mu\nu} + \GBc(\gamma'\kappa_d)^{2}d(d-3)e^{-\gamma'\kappa_d\varphi}\pd_\mu\varphi\pd_\nu\varphi,
\ee
which is allowed by the on-shell nature of string scattering amplitudes.
Then, we find that the action \eqref{eqn:GBfalseeinstein} in this new frame becomes
\begin{align}
I[g,\varphi] = \frac{1}{2\dgravc}\int d^dx\sqrt{-g}\biggl[ &R - \dgravc(\pd\varphi)^2
                                                                                    + {\GBc}e^{-\gamma'\kappa_d\varphi}\biggl\{ \GBinv(g)
                                                                                    + {\half}(\gamma'\kappa_d)^3(d-1)_3\nabla^2\varphi(\pd\varphi)^2 \notag\\
                                                                                   &+ \frac{1}{16}(\gamma'\kappa_d)^{4}(d^2-3d+4)(d-2)(d-3)\left\{(\pd\varphi)^2\right\}^2 \biggr\}\biggr]
                                                                                    + O(\regge^2),
                                                                                    \label{eqn:GBeinstein}
\end{align}
where the superscript ``new'' is omitted for simplicity. As is desired, a dilaton does minimally couple to gravity up to the first order in the curvature.
Thus we conclude that this action \eqref{eqn:GBeinstein} describes the DEGB gravity written in Einstein frame.

The kinetic term for a dilaton in this action takes the form
\begin{gather}
-{\half}f_d(\varphi)(\pd\varphi)^2; \notag \\
f_d(\varphi) \defeq
   1 - \frac{\gamma'^3\kappa_d\regge}{16}(d-2)_{3}e^{-\gamma'\kappa_d\varphi}
                            \left( (d - 1)\nabla^2\varphi + \frac{1}{8}\gamma'\kappa_{d}(d^{2} - 3d + 4)(\pd\varphi)^{2} \right), \label{eqn:kt}
\end{gather}
i.e. the noncanonical one.
As is stated above, in the regime of low energy effective theory the Regge slope is extremely small,
so we could expect that the following relation is generic:
\nbe
f_d(\varphi) > 0.
\footnote{Actually, the function $f_d$ is not positive definite, so this condition could be violated, depending on a dilaton dynamics whose current status
is `in confusion,' because its scalar potential is yet-to-be discovered. Thus one may use it as the condition which should be satisfied by a dilaton.}
\nee
Then, there exists the redefinition of a dilaton $\varphi\rightarrow\psi = \psi(\varphi)$ such that
the kinetic term \eqref{eqn:kt} can be brought to a canonical form.
Therefore, we observe that the DEGB gravity in the Einstein frame is described by the action
\be
I[g,\psi] = \frac{1}{2\dgravc}\int d^dx\sqrt{-g}\left( R - \dgravc(\pd\psi)^2 + {\GBc}e^{-\gamma'\kappa_d\psi}\GBinv(g) \right).
\label{eqn:GBfinal}
\ee
It takes a relatively simple form, like the original string frame action.

\section{Conclusions}
In this note we analyzed the properties of the DEGB gravity, arising as the low energy effective theory of heterotic string,
and proposed the transition to the Einstein frame action.
It receives a modification from the viewpoint of EEP, coming from the dynamical nature of the dilaton.
It is demonstrated that, by making use of the smallness of the Regge slope $\regge$ in field theory regime,
the kinetic term of a dilaton can be generically brought to a canonical form.
It is observed that the DEGB gravity action in the Einstein frame has a clear form in terms of the redefined dilaton,
leading to our final result \eqref{eqn:GBfinal}.

We observed that the definition of the Einstein frame depends on the order of string perturbation theory.
From the viewpoint of perturbative strings, the modification of a gravitational action is given by the power series expansion
in the Regge slope $\regge$ (so infinitely many $\regge$-corrections exist).
Since the DEGB gravity is just the leading order correction to GR, our results are supposed to receive corrections
when more higher-order terms in $\regge$ are taken into account.

\vskip 10pt

{\centering{\subsection*{Acknowledgements}}}
The author is very grateful to S. V. Ketov for useful discussions and valuable comments on the draft of this paper,
as well as S. Deser, N. E. Mavromatos, and S. D. Odintsov for correspondence.

    \end{document}